\documentclass[aps,showpacs]{revtex4}%
\usepackage{amsfonts}%
\usepackage{amsmath}%
\usepackage{amssymb}%
\usepackage{graphicx}

\begin{document}
\title{Complementarity in a macroscopic observation}
\author{De-Zhong Cao,$^{1,2}$ Jun Xiong,$^{1}$ Hua Tang,$^{1}$ Lu-Fang Lin,$^{1}$
Su-Heng Zhang,$^{1}$ and Kaige Wang$^{1,}$\footnote{Author to whom
correspondence should be addressed. Electronic address:
wangkg@bnu.edu.cn}}
\affiliation{$^{1}$Department of Physics,
Applied Optics Beijing Area Major Laboratory, Beijing Normal University, Beijing 100875, China\\
$^{2}$Department of Physics, Yantai University, Yantai 264005, China}
\begin{abstract}
Complementarity is usually considered as a phenomenon of microscopic systems.
In this paper we report an experimental observation of complementarity in the
correlated double-slit interference with a pseudothermal light source. The
thermal light beam is divided into test and reference beams which are
correlated with each other. The double-slit is set in the test arm, and the
interference pattern can be observed in the intensity correlation between the
two arms. The experimental results show that the disappearance of interference
fringe depends on whether which-path information is gained through the
reference arm. The experiment therefore witnesses the complementarity
occurring in a macroscopic system.

\end{abstract}
 \maketitle

Since the principle of complementarity was originally proposed in the dialogue
between Bohr and Einstein it has drawn much attention and aroused interesting
debates in the past
years\cite{scly1,mandel,sto,scly2,sto1,chiao92,zeilinger95,chap,durr,sch,shih00,tt,ber,monken02,and,gogo05}%
. The principle states that simultaneous observation of wave and particle
behavior of a microscopic system is prohibited. In a Young's double-slit
interference experiment, for example, one can never simultaneously obtain
exact knowledge of the photon trajectory (particle behavior) and the
interference fringes (wave behavior). To be precise, if the photon trajectory
is definitely known, no interference pattern can be observed. Accordingly, if
an interference pattern is recorded, the photon path cannot be distinguished.

As quantum phenomena occurring in the microscopic realm, the original gedanken
experiments and the following experimental implementions of complementarity
involved individual microscopic particles or
photons\cite{mandel,chiao92,zeilinger95,chap,durr,sch,shih00,tt,ber,monken02,and,gogo05}%
. In this paper we report the observation of complementarity between
interference fringe and which-path knowledge in a correlated double-slit
interference experiment with thermal light. We describe this as a
\textit{macroscopic observation} since the source used in our experiment is a
pseudothermal light beam and measurement is performed by ordinary optical
intensity detection, rather than single-photon detection. To avoid direct
measurement, we utilize the spatial correlation of thermal light to set up a
telltale apparatus and gain which-path information.

Recent studies have shown that a thermal light source can play a role similar
to that of a two-photon entangled source in "ghost" imaging, "ghost"
interference and subwavelength
interference\cite{lugiato04,cheng04,cao05,lugiato05,dz,wang04,cai04,xiong05,zhai05,shih06}.
When a thermal light beam random in its transverse wavevector illuminates a
double-slit, the interference pattern does not appear in the intensity
distribution but can be extracted in the intensity
correlation\cite{wang04,cai04,xiong05,zhai05}. On the other hand, the thermal
light can also mimic entangled photon pairs in performing ghost
imaging\cite{lugiato04,cheng04,cao05,lugiato05,dz,shih06}. By analyzing different correlation
features between entangled photon pairs and thermal light, Ref. \cite{cao05}
first pointed out that thermal light can, moreover, exhibit ghost imaging
without using any lenses. That is, the beamsplitter which divides the thermal
light into two beams acts as a phase conjugate mirror and a conjugate image
can be formed at the symmetric position of the object with respect to the
beamsplitter. It thus establishes point-to-point correspondence between the
object and image. As a result, ghost imaging manifests particle-like behavior:
if one photon illuminates a point on the object, the other photon must arrive
at the corresponding position of the image. The path information of photons in
one arm can be extracted by knowledge of the other arm of the beamsplitter.

In this paper we discuss two experimental schemes. The first is depicted in
Fig. 1. A pseudothermal light beam, which is formed by a He-Ne laser beam
projected onto a rotating ground glass, is divided by two 50/50 beamsplitters
(BS1 and BS2) into three beams: one test beam which illuminates a double-slit,
and two reference beams which propagate freely. Three charge-coupled device
(CCD) cameras are used to register the beam intensity in each arm: CCD1
detects the intensity of the beam passing through the double-slit in the test
arm while CCD2 and CCD3 register the intensity distributions in the two
reference arms. We place CCD2 at the symmetric position of the double-slit
with respect to BS2 and CCD3 in the far field, thus, in the intensity
correlation with CCD1, the former may show ghost imaging and the latter ghost
interference. This scheme is similar to that proposed in Ref. \cite{shih04}
where the authors reported experimentally that ghost imaging and ghost
interference can be implemented simultaneously with an entangled two-photon
state, and claimed this would be a distinction between classically correlated
and quantum entangled systems.

Figure 2 shows the experimental results. Note that CCD2 and CCD3 register the
intensity distributions $I_{2}(x_{2})$ and $I_{3}(x_{3})$ across the reference
arms, respectively, while CCD1 detects the intensity $I_{1}(x_{1})$ at a fixed
position $x_{1}=0$ in the test arm. In Fig. 2a the normalized intensity
correlation between CCD1 and CCD2 $\langle I_{1}(0)I_{2}(x_{2})\rangle
/(\langle I_{1}(0)\rangle\langle I_{2}(x_{2})\rangle)$ shows a conjugate image
of the double-slit, and in Fig. 2b, the correlation $\langle I_{1}%
(0)I_{3}(x_{3})\rangle/(\langle I_{1}(0)\rangle\langle I_{3}(x_{3})\rangle)$
between CCD1 and CCD3 exhibits an interference pattern. This demonstrates that
ghost imaging and ghost interference can be simultaneously observed with a
pseudothermal light source. Nevertheless, the experimental results do not
display any complementarity between interference and imaging. In this
experiment the photons which participate in ghost imaging and have been
detected by CCD2 are never involved in the interference, and vice versa.

We now propose another scheme to monitor the which-path information in the
correlated double-slit interference. As shown in Fig. 3, the double-slit is
placed in the test arm, but now a single-slit is inserted into the reference
arm at the position that is exactly symmetric relative to the beamsplitter and
coincides with one slit of the double-slit in the test arm. This single-slit
plays the telltale role as a measuring apparatus due to the point-to-point
correspondence between object and image. That is, the which-path information
of the double-slit interference in the test arm can be obtained through the
single-slit in the reference arm.

To demonstrate the ghost interference, CCD1 and CCD2 register the intensity
$I_{1}(x_{1})$ after the double-slit and $I_{2}(x_{2})$ after the single-slit,
respectively. Figure 4a shows the normalized intensity correlation $\langle
I_{1}(x_{1})I_{2}(x_{2})\rangle/(\langle I_{1}(x_{1})\rangle\langle
I_{2}(x_{2})\rangle)$, where the left plot indicates the correlation
distribution measured by scanning the position $x_{1}$ in the reference arm
for a fixed position $x_{2}=0$ in the test arm, and vice versa for the right
plot. The two correlation curves do not exhibit any interference patterns.

One might think that the disappearance of interference in the above scheme is
due to the disturbance of an aperture inserted into the reference arm. In
order to confirm the effectiveness of the present telltale apparatus, in Fig.
3, we next replace the single-slit in the reference arm with a double-slit,
which is exactly the same as in the test arm. In this case the which-path
information is completely erased. The correlation measurement results are
shown in Fig. 4c, where we can see that the correlated interference fringes
reappear with a higher visibility than in Fig. 2 of the first scheme.
Furthermore, when we partly cover one slit of the double-slit in the reference
arm, incomplete which-path information is gained. Figure 4b shows that
although the interference fringes are still observed they have a lower
visibility. These experimental results clearly demonstrate the complementarity
in correlated double-slit interference with thermal light: any attempt to
extract path information from the reference arm shall degrade the interference
fringe visibility.

Our experimental results can be explained by considering the spatial
correlation properties of thermal light. The thermal light source used in the
experiment is described by the field distribution $E(\mathbf{x}%
,z,t)=\int\widetilde{E}(\mathbf{q})\exp[i\mathbf{q}\cdot\mathbf{x}%
]d\mathbf{q}\times\exp[i(kz-\omega t)]$, where $z$ is the propagation
direction and $\mathbf{x}$ and $\mathbf{q}$ the transverse position and
wavevector, respectively. According to the Wiener-Khintchine theorem, the
first-order spectral correlation of the stochastic field $\widetilde
{E}(\mathbf{q})$ can be written as $\langle\widetilde{E}^{*}(\mathbf{q}%
)\widetilde{E}(\mathbf{q}^{\prime})\rangle=W(\mathbf{q})\delta(\mathbf{q}%
-\mathbf{q}^{\prime})$, where $W(\mathbf{q})$ is the power spectrum of the
spatial frequency. By Fourier transform the first-order spatial correlation of
the field is given by $\langle E^{*}(\mathbf{x})E(\mathbf{x}^{\prime}%
)\rangle=\widetilde{W}(\mathbf{x}-\mathbf{x}^{\prime})$, where the correlation
function $\widetilde{W}(\mathbf{x})=\frac1{2\pi}\int W(\mathbf{q}%
)\exp(i\mathbf{q}\cdot\mathbf{x})d\mathbf{q}$. For thermal light, the
second-order spatial correlation satisfies
\begin{equation}
\langle E^{*}(\mathbf{x}_{1})E^{*}(\mathbf{x}_{2})E(\mathbf{x}_{2}^{\prime
})E(\mathbf{x}_{1}^{\prime})\rangle=\langle E^{*}(\mathbf{x}_{1}%
)E(\mathbf{x}_{1}^{\prime})\rangle\langle E^{*}(\mathbf{x}_{2})E(\mathbf{x}%
_{2}^{\prime})\rangle+\langle E^{*}(\mathbf{x}_{1})E(\mathbf{x}_{2}^{\prime
})\rangle\langle E^{*}(\mathbf{x}_{2})E(\mathbf{x}_{1}^{\prime})\rangle
.\label{3}%
\end{equation}

We now consider the propagation of the field in the test and reference arms.
Let $E(\mathbf{x})$ be a source field, so the outgoing field is obtained as
\begin{equation}
E_{j}(\mathbf{x})=\int h_{j}(\mathbf{x},\mathbf{x}^{\prime})E(\mathbf{x}%
^{\prime})d\mathbf{x}^{\prime},\qquad\ (j=1,2)\label{5}%
\end{equation}
where $h_{j}(\mathbf{x},\mathbf{x}_{0})$ $(j=1,2)$ is the transfer function
describing the test and reference systems, designated by indices 1 and 2,
respectively. The intensity correlation between the test and reference arms
can be calculated from
\begin{align}
\left\langle I_{1}(\mathbf{x}_{1})I_{2}(\mathbf{x}_{2})\right\rangle  &
=\langle E_{1}^{*}(\mathbf{x}_{1})E_{2}^{*}(\mathbf{x}_{2})E_{2}%
(\mathbf{x}_{2})E_{1}(\mathbf{x}_{1})\rangle\nonumber\\
& =\langle I_{1}(\mathbf{x}_{1})\rangle\langle I_{2}(\mathbf{x}_{2}%
)\rangle+|\langle E_{1}^{*}(\mathbf{x}_{1})E_{2}(\mathbf{x}_{2})\rangle
|^{2},\label{4}%
\end{align}
where
\begin{subequations}
\label{6}%
\begin{align}
\left\langle I_{j}(\mathbf{x})\right\rangle  & =\frac1{2\pi}\int h_{j}%
^{*}(\mathbf{x},\mathbf{x}_{0}^{\prime})h_{j}(\mathbf{x},\mathbf{x}%
_{0})\widetilde{W}(\mathbf{x}_{0}^{\prime}-\mathbf{x}_{0})d\mathbf{x}%
_{0}^{\prime}d\mathbf{x}_{0},\qquad(j=1,2)\label{6a}\\
\left\langle E_{1}^{*}(\mathbf{x}_{1})E_{2}(\mathbf{x}_{2})\right\rangle  &
=\frac1{2\pi}\int h_{1}^{*}(\mathbf{x}_{1},\mathbf{x}_{0}^{\prime}%
)h_{2}(\mathbf{x}_{2},\mathbf{x}_{0})\widetilde{W}(\mathbf{x}_{0}^{\prime
}-\mathbf{x}_{0})d\mathbf{x}_{0}^{\prime}d\mathbf{x}_{0}.\label{6b}%
\end{align}
In Eq. (\ref{4}) the first term contributes a background, and the second term
may contain the coherence information, which can be extracted by intensity
correlation measurements.

For simplicity we consider the one-dimensional case. The transfer function for
free travel over a distance $z$ is given by
\end{subequations}
\begin{equation}
h(x,x_{0})=\sqrt{\frac k{2\pi iz}}\exp(ikz)\exp\left[  ik\frac{(x-x_{0})^{2}%
}{2z}\right]  .\label{7}%
\end{equation}
Thus the field correlation between the test and reference arms at the same
distance $z$ is obtained as
\begin{align}
\left\langle E_{1}^{*}(x_{1})E_{2}(x_{2})\right\rangle  & =\frac
k{(2\pi)^{3/2}z}\exp\left[  i\frac k{2z}(x_{2}^{2}-x_{1}^{2})\right]
\nonumber\\
& \times\int\exp\left[  i\frac k{2z}(x_{0}^{2}-x_{0}^{\prime2}-2x_{0}%
x_{2}+2x_{0}^{\prime}x_{1})\right]  \widetilde{W}(x_{0}^{\prime}-x_{0}%
)dx_{0}dx_{0}^{\prime}.\label{8}%
\end{align}
In the broadband limit of the spectrum, $\widetilde{W}(x)\rightarrow\sqrt
{2\pi}W_{0}\delta(x)$, Eq. (\ref{8}) becomes
\begin{equation}
\left\langle E_{1}^{*}(x_{1})E_{2}(x_{2})\right\rangle =W_{0}\delta
(x_{1}-x_{2})\label{9}%
\end{equation}
which shows a point-to-point correspondence of the field amplitudes between
the two arms. This feature is also reflected in the intensity correlation of
Eq. (\ref{4}). In the quantum regime, however, Eq. (\ref{4}) describes a
two-photon coincidence probability. Therefore, knowledge of a photon's
position in one arm implies knowledge of the position of the correlated photon
in the other arm. This provides telltale information about the photon paths
without disturbing the interference system. Moreover, Eqs. (\ref{8}) and
(\ref{9}) are the origin of correlated imaging without the use of lenses.

When an object of transmission $T(x)$ is inserted in the test arm, the
transfer function is written as
\begin{equation}
h(x,x_{0})=\frac k{i\sqrt{2\pi z_{0}z}}\exp[ik(z_{0}+z)]\exp\left[  i\frac
k2\left(  \frac{x^{2}}z+\frac{x_{0}^{2}}{z_{0}}\right)  \right]  \widetilde
{T}\left[  k\left(  \frac xz+\frac{x_{0}}{z_{0}}\right)  \right]  ,\label{10}%
\end{equation}
where $\widetilde{T}$ is the Fourier transform of $T$; $z_{0}$ and $z$ are the
distances from source to object and from object to detector, respectively. In
the paraxial approximation, for a double-slit $D(x)$ and a single-slit $S(x)$,
$\widetilde{T}(q)$ is replaced by
\begin{subequations}
\label{11}%
\begin{align}
\widetilde{D}(q)  & =(2b/\sqrt{2\pi})\text{sinc}(qb/2)\cos(qd/2),\label{11a}\\
\widetilde{S}(q)  & =(b/\sqrt{2\pi})\text{sinc}(qb/2)\exp(-iqd/2),\label{11b}%
\end{align}
respectively, where $b$ is the slit width and $d$ is the distance between the
centers of two slits.

In the broadband limit, by using Eqs. (\ref{6}), (\ref{10}) and (\ref{11}) we
obtain the analytical solution of the intensity correlation for the scheme of
Fig. 3. When a double-slit and a single-slit are inserted into the reference
arm, we obtain
\end{subequations}
\begin{subequations}
\label{12}%
\begin{align}
\left\langle I_{1}(x_{1})I_{2}(x_{2})\right\rangle  & =\frac{k^{2}W_{0}^{2}%
}{2\pi z^{2}}\left\{  \widetilde{D}^{2}(0)+\widetilde{D}^{2}\left[  \frac
kz(x_{1}-x_{2})\right]  \right\}  ,\label{12a}\\
\left\langle I_{1}(x_{1})I_{2}(x_{2})\right\rangle  & =\frac{k^{2}W_{0}^{2}%
}{2\pi z^{2}}\left\{  \widetilde{D}(0)\widetilde{S}(0)+\left|  \widetilde
{S}\left[  \frac kz(x_{1}-x_{2})\right]  \right|  ^{2}\right\}  ,\label{12b}%
\end{align}
respectively, where $z$ is the distance between the slit and detector.
Equation (\ref{12a}) gives interference fringes with a visibility of 33.3\%,
which is the maximum obtainable in correlated double-slit interference (see
Fig. 4c). Instead of interference fringes, however, Eq. (\ref{12b}) produces a
diffraction pattern, as seen in Fig. 4a.

As for the incomplete double-slit in the reference arm where the width of one
slit is reduced to one quarter, its Fourier function is given by
\end{subequations}
\begin{equation}
\widetilde{Q}(q)=\widetilde{S}(q)+\frac b{4\sqrt{2\pi}}\text{sinc}(\frac
{qb}8)\exp[iq(\frac d2-\frac{3b}8)].\label{13}%
\end{equation}
The correlation function is calculated to be
\begin{equation}
\left\langle I_{1}(x_{1})I_{2}(x_{2})\right\rangle =\frac{k^{2}W_{0}^{2}}{2\pi
z^{2}}\left\{  \widetilde{D}(0)\widetilde{Q}(0)+\left|  \widetilde{Q}\left[
\frac kz(x_{1}-x_{2})\right]  \right|  ^{2}\right\}  ,\label{14}%
\end{equation}
and the interference fringes have a visibility of 23.8\% (see Fig. 4b) lower
than that for the complete double-slit. The numerical simulations agree well
with the experimental data.

In summary, the correlated double-slit interference phenomenon of thermal
light manifests wave behavior. However, the correlated imaging of thermal
light exhibits particle-like behavior since it indicates the position-position
correspondence between two beams. We have demonstrated that, in the correlated
double-slit interference, any attempt to acquire path-information from the
reference system will destroy the interference. The present experiment can be
considered as a macroscopic observation of the complementarity, and may
promote our knowledge of the quantum world.

The authors thank L. A. Wu for helpful discussions. This research was
supported by the National Fundamental Research Program of China Project No.
2001CB309310, and the National Natural Science Foundation of China, Project
No. 10574015. One of the authors D. -Z. Cao is grateful to the support by
National Natural Science Foundation of China, Project No. 10547136.

Figure captions:

Figure 1. Schematic of experimental setup for simultaneously observing ghost
imaging and ghost interference with a pseudo-thermal light beam. The source
beam is split into three by two beamsplitters (BS1 and BS2), and their
intensities can be adjusted by polarizers P$_1$, P$_2$, and P$_3$, and then registered
by three charge coupled device (CCD) cameras (MINTRON:MTV-1881EX). The
double-slit has a slit width $b=85\mu m$ and slit-center separation $d=330\mu
m$. The distances from the ground glass to the double-slit, CCD1, CCD2, and
CCD3 are 10, 30, 10, and 44$cm$, respectively.

Figure 2. Experimental data of intensity distributions (solid squares) and
normalized intensity correlations (circles) registered by (a) CCD2 and (b)
CCD3. In the measurements, CCD1 detects a fixed position in the test arm while
CCD2 and CCD3 register the intensity distributions across the reference arms.
Numerical simulations are shown by solid lines.

Figure 3. Schematic of experimental setup for observing the complementarity
with pseudo-thermal light. The reflected (test) arm contains a double-slit,
while the transmitted (reference) arm may contain a single-slit, a
double-slit, or an incomplete-slit. The distances from the ground glass to the
double-slit, the reference arm slits, CCD1, and CCD2 are 2, 2, 42 and 42$cm$, respectively.

Figure 4. Experimental data of intensity distributions (solid squares) and
normalized intensity correlations (circles) for (a) a single-slit, (b) an
incomplete double-slit, and (c) a double-slit placed in the reference arm. The
left column figures show the results when CCD1 detects a fixed position in the
test arm while CCD2 registers the intensity distribution across the reference
arm, and vice versa for the right column. Numerical simulations are shown by
solid lines.

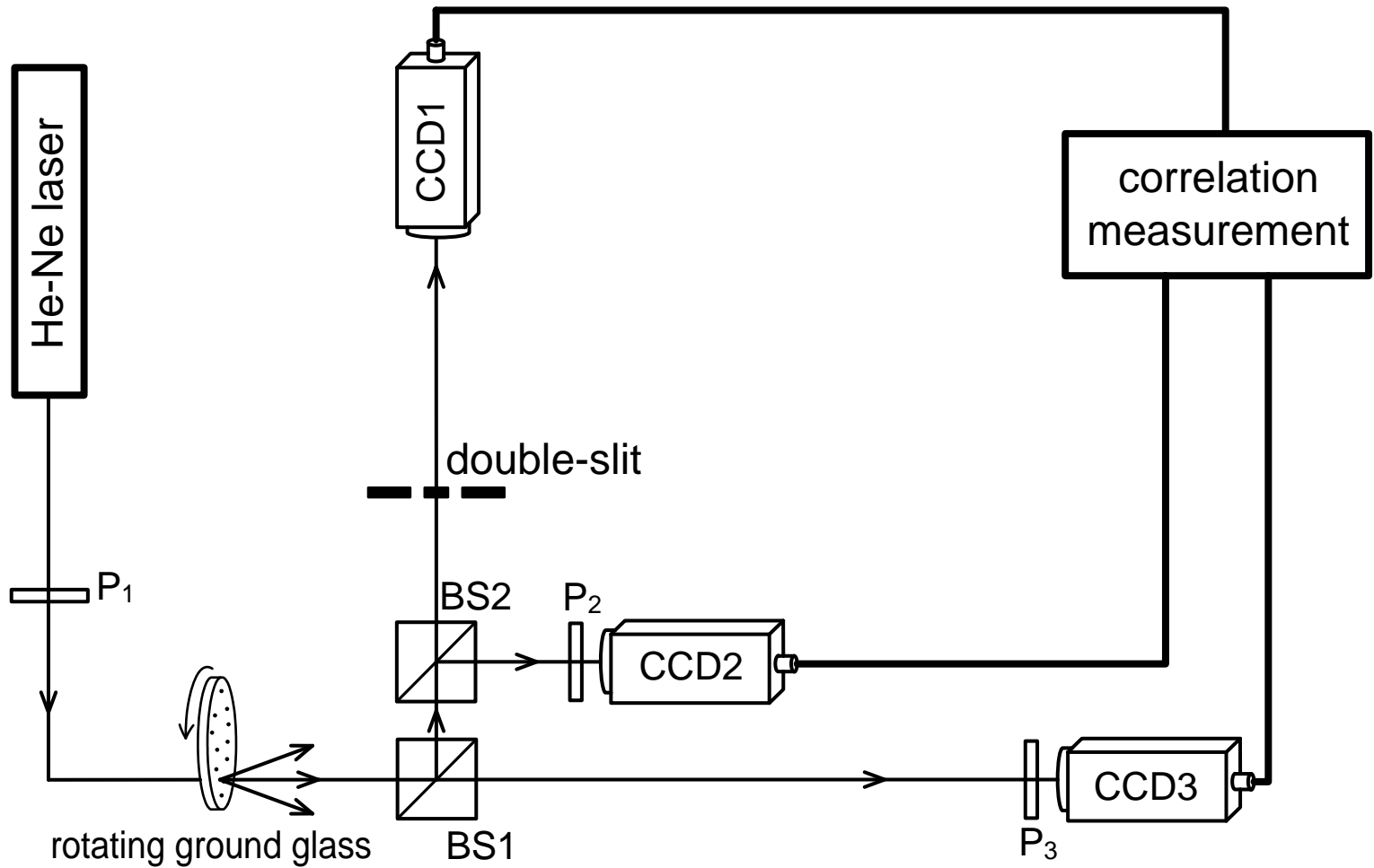
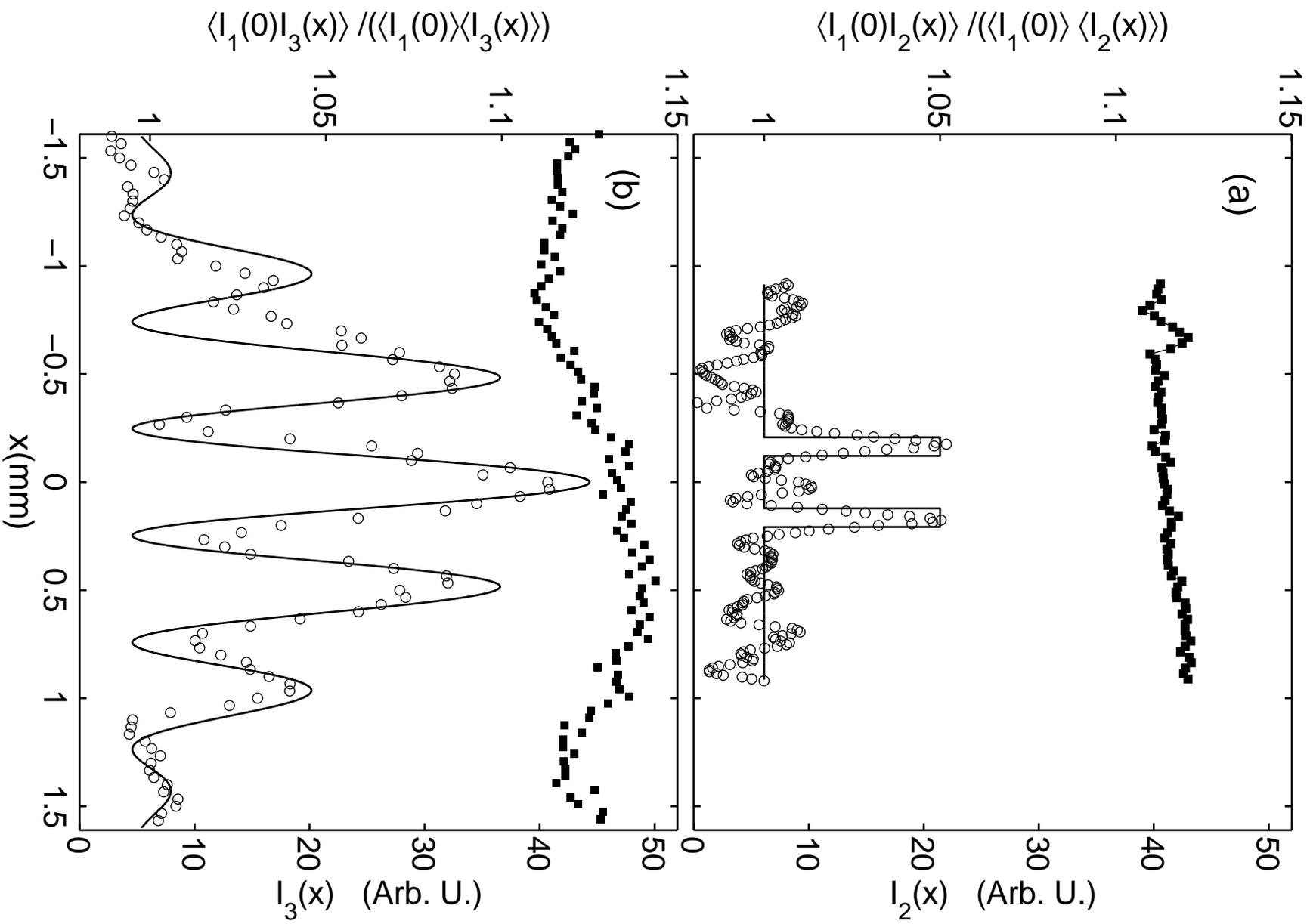
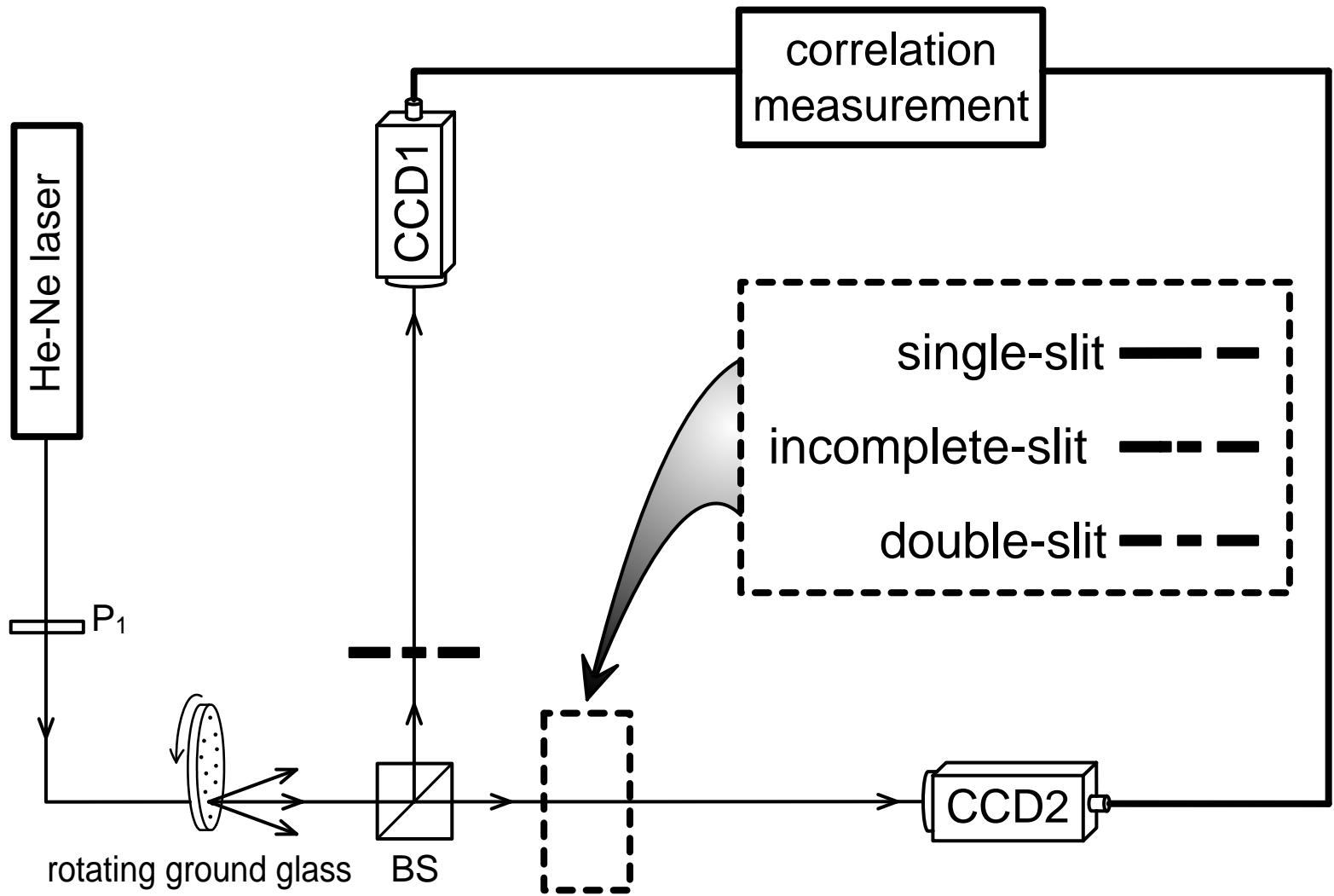
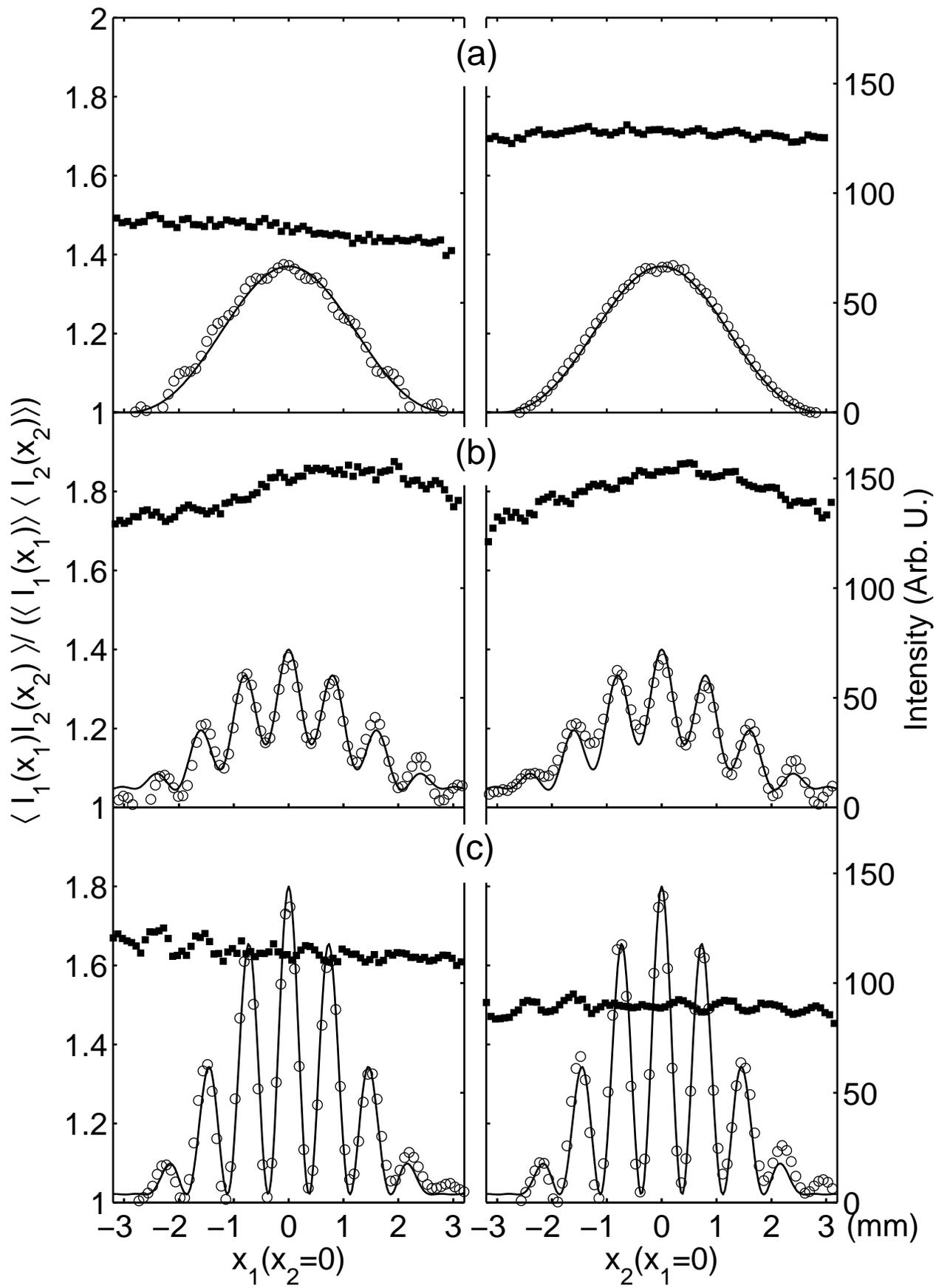
\end{document}